\documentclass[10pt,aps,prd,twocolumn,nofootinbib,sort&compress,superscriptaddress]{revtex4-1}

\usepackage{amssymb,amsmath,accents}
\usepackage{subcaption,verbatim}
\usepackage{graphicx}
\usepackage{color,units}
\usepackage[table,xcdraw,dvipsnames]{xcolor}
\usepackage{hyperref}
\usepackage{soul} 
\usepackage{listings}
\usepackage{ragged2e}
\usepackage{aas_macros}
\usepackage{placeins}
\usepackage{float}
\usepackage{orcidlink}
\usepackage{aas_macros}
\usepackage{natbib}

\usepackage[normalem]{ulem}
\hypersetup{
    bookmarks=true,         
    unicode=false,          
    pdftoolbar=true,        
    pdfmenubar=true,        
    pdffitwindow=false,     
    pdfstartview={FitH},    
    pdfauthor={Boris Goncharov},     
    colorlinks=true,       
    linkcolor=blue,          
    citecolor=red,        
}
\graphicspath{{figures/}}

\begin{document}

\definecolor{dkgreen}{rgb}{0,0.6,0}
\definecolor{gray}{rgb}{0.5,0.5,0.5}
\definecolor{mauve}{rgb}{0.58,0,0.82}

\lstset{frame=tb,
  	language=Matlab,
  	aboveskip=3mm,
  	belowskip=3mm,
  	showstringspaces=false,
  	columns=flexible,
  	basicstyle={\small\ttfamily},
  	numbers=none,
  	numberstyle=\tiny\color{gray},
 	keywordstyle=\color{blue},
	commentstyle=\color{dkgreen},
  	stringstyle=\color{mauve},
  	breaklines=true,
  	breakatwhitespace=true,
  	tabsize=3
}

\title{The Targeted Standard Siren Cosmology with Pulsar Timing Arrays} 

\author{Shubhit Sardana\, \orcidlink{0009-0007-2913-7704}}
\affiliation{Department of Physics, Indian Institute of Science Education and Research Bhopal, 462066 Bhopal, India}
\email{sardanashubhit@gmail.com}

\author{Boris Goncharov\, \orcidlink{0000-0003-3189-5807}}
\affiliation{Max Planck Institute for Gravitational Physics (Albert Einstein Institute), 30167 Hannover, Germany}
\affiliation{Leibniz Universität Hannover, 30167 Hannover, Germany}

\author{Jacob Cardinal Tremblay\,\orcidlink{0000-0001-9852-6825}}
\affiliation{Max Planck Institute for Gravitational Physics (Albert Einstein Institute), 30167 Hannover, Germany}
\affiliation{Leibniz Universität Hannover, 30167 Hannover, Germany}

\date{\today}

\begin{abstract}

The sky localisation of about $10$ to $100~\text{deg}^2$, which is expected to be achieved in all-sky blind searches for gravitational waves from supermassive black hole binaries (SMBHBs) with Pulsar Timing Array (PTA) experiments, has long been posed as a prohibitive factor in utilising these sources as standard sirens for precision cosmology. 
We propose a solution to this problem, which makes use of targeted searches rather than all-sky blind searches for SMBHBs. 
Using our simulated data informed by current PTA observations, we show that the Chinese Pulsar Timing Array (CPTA) alone could infer the Hubble constant with a precision of 2~km/s/Mpc. 
Such precision in an independent cosmological probe could provide decisive support in the resolution of the Hubble tension. 
We demonstrate the application of our method to several simultaneously observed SMBHBs, as well as the method's robustness against confusion between the host galaxies of SMBHB sources in realistic observing scenarios. 

\end{abstract}

\maketitle

Observations of gravitational waves (GWs) from inspiralling compact binaries provide measurements of cosmological luminosity distances. 
When electromagnetic counterparts of these signals are identified, compact binaries turn into the so-called standard sirens.
Such standard sirens allow a direct measurement of the luminosity distance without relying on additional methods such as the distance ladder~\cite{Schutz1986}. 
By using the luminosity distance information from GWs and the redshift information from the electromagnetic counterpart, a measurement of the Hubble constant $H_0$ becomes possible. 
A successful identification of the host galaxy for a binary neutron star system inspiralling due to the GW emission, GW170817~\citep{gw170817}, has enabled a measurement of $H_0 = 70.0^{+12.0}_{-8.0}\ [\text{km s}^{-1}\ \text{Mpc}^{-1}]$~\cite{gw170817_h0}. 
Although the precision of this measurement is about an order of magnitude lower than for state-of-the-art cosmological experiments like Planck ($H_0 = 67.4\pm0.5\ [\text{km s}^{-1}\ \text{Mpc}^{-1}]$~\citep{planck_2021}) and SH0ES ($H_0 = 73.04\pm1.04\ [\text{km s}^{-1}\ \text{Mpc}^{-1}]$~\citep{RiessYuan2022}), a 2-\% measurement of $H_0$ may be possible in the near future~\cite{ChenFishbach2018}. 
As an independent technique, standard sirens could thus contribute significantly to precision cosmology and even assist in resolving the aforementioned discrepancy between $H_0$ estimates of SH0ES and Planck, known as the Hubble tension~\cite{DiValentinoMena2021, VerdeSchoneberg2024}. 

Like other gravitational wave experiments, Pulsar Timing Arrays (PTAs) could also be used to identify standard sirens. 
PTAs have a primary goal of detecting gravitational waves at nanohertz frequencies through decade-long monitoring of pulse arrival times from galactic millisecond pulsars. 
Independent PTA collaborations have found multiple lines of evidence for the stochastic gravitational-wave background (GWB)~\citep{GoncharovShannon2021, ng12_gwb, ChenCaballero2021, ipta_dr2_gwb, ng15_gwb, epta_dr2_gwb, ReardonZic2023}. 
If the GWB is formed by a superposition of GWs from Supermassive Black Hole Binaries (SMBHBs), as expected, SMBHBs should become individually resolvable. 
All-sky searches for continuous gravitational waves (CGWs) from single SMBHBs are underway~\citep{ng15_cw, ZhaoChen2025, epta_dr2_cw, ipta_dr2_cw}. 
Identification of the host galaxy for an SMBHB from which a CGW is detected could enable a standard siren observation with a PTA. 

In all-sky searches for CGWs with PTAs, the identification of the host galaxy of the signal required for cosmological distance measurements is challenging. 
Expecting sky position uncertainties of tens to hundreds of square degrees~\cite{PetrovTaylor2024,PetrovSchult2025,ZhuWen2016,CharisiTaylor2024}, SMBHBs at the early inspiral stage are not expected to produce bright electromagnetic transients, which would point to the specific host galaxy. 
Thus, the literature on standard sirens with PTAs is scarce. 
A solution proposed by D'Orazio and Loeb~\cite{D'OrazioLoeb2021} is based on the additional measurement of the gravitational-wave comoving distance.
It is obtained by probing the GW wavefront curvature, a measurement which is unique to PTAs as GW detectors. 
Simultaneously with the luminosity distance, the comoving distance measurement allows estimation of the source redshift and thus the Hubble constant.
This allows constraining $H_0$ to less than a $30\%$ level. 
The difficulty with this result is that the reported precision is still below that obtained with GW170817, and even this requires high-precision pulsar distance measurements up to a CGW wavelength. 
A technique known as ``dark sirens'' is another possible solution that utilises unlocalised GW signals for cosmological inference by considering the distribution of possible host galaxies~\cite{SoaresSantosPalmese2019}. 
A recent study finds that a PTA with $100$ pulsars timed at $20$ ns precision may achieve a per cent measurement of the Hubble constant with $150$ dark sirens~\cite{XiaoShao2025}. 
A similar study~\cite{WangShao2025} considers both up to $56$ dark sirens and up to $53$ localized CGWs as conventional standard sirens (bright sirens), achieving a comparable precision on $H_0$ with a similar toy analysis. 
The authors find that the analysis can further be used to infer the dark energy parameter $w$. 
However, Ref.~\cite{WangShao2025} only considers a detection of CGWs, not a localisation procedure.
The authors assume that all their simulated SMBHBs are real, and point out that their sky position uncertainties of less than $1~\deg^2$--which are one order of magnitude less than the most optimistic predictions in PTA simulations--are sufficient to distinguish catalogue sources from each other. 
Moreover, the limitations of Refs.~\cite{XiaoShao2025,WangShao2025} include the use of the Fisher matrix approximation of the PTA likelihood without the full consideration of time-correlated processes in PTA data, including the GWB and noise. 

An alternative, rather unexplored approach to standard sirens with PTAs makes use of targeted searches for SMBHBs instead of all-sky searches. 
Targeted CGW searches with PTAs follow up on Active Galactic Nuclei (AGN) that are expected to host SMBHBs.
In targeted searches, uninformative priors on CGW parameters such as the GW frequency $f$, the chirp mass $\mathcal{M}$, the sky position $\hat{\Omega}$, and the luminosity distance $D_\text{L}$, are replaced by informative priors based on the results of SMBHB parameter estimation with the electromagnetic radiation data from AGN. 
Because the AGN's sky position is known, targeted CGW searches make it unnecessary to consider multiple galaxies, perform detailed electromagnetic follow-up, or achieve exceptional sky-position uncertainty. 
The idea to employ targeted searches to infer $H_0$ has been proposed in the targeted search for the SMBHB in elliptical galaxy 3C~66B with the Parkes Pulsar Timing Array (PPTA)~\cite{CardinalTremblayGoncharov2026}. 
The authors have used an unconstrained posterior on the $D_\text{L}$ of 3C~66B and the knowledge of its redshift $z$ to demonstrate constraints on $H_0$, conditioned on the validity of the target. 
In Ref.~\cite{WangShao2025}, the authors have also hinted at this possibility because their bright sirens are SMBHB candidates from catalogues like the Catalina Real-Time Transient Survey~\cite{GrahamDjorgovski2015}. 
This catalogue reports periodicity in light curves attributable to SMBHBs, which have been considered in the recent targeted search for CGWs by the North American Nanohertz Observatory for Gravitational Waves (NANOGrav)~\cite{ng15_cw_target}.
It is worth noting, however, that some, if not many, SMBHB candidates may be false positives~\cite{El-BadryHogg2026,VaughanUttley2016}. 
A full parameter estimation study with realistic simulated binaries and noise, as well as the proper assessment of the source confusion in the SMBHB population, is necessary to determine the prospects of the targeted standard siren cosmology with PTAs. 

In this work, we assess the feasibility of the standard sirens based on targeted CGW searches with PTAs. 
To achieve this, perform Bayesian parameter estimation for CGW parameters, noise parameters, as well as the strain power-law amplitude and spectral index of the GWB in the simulated data. 
The Hubble constant $H_0$ is parameterised in terms of the CGW luminosity distance and target SMBHB redshifts. 
Compared to Ref.~\cite{CardinalTremblayGoncharov2026}, we perform inference simultaneously with up to 3 CGW sources, two of which have been recently associated with marginally significant outliers in the targeted search for CGWs by NANOGrav~\cite{ng15_cw_target}. 
Furthermore, through simulating a synthetic population of SMBHBs, we quantify the extent of the CGW source confusion and misspecification, reporting whether it is a limiting factor for targeted standard sirens.

\section*{\label{sec:results} Results}

We obtain the results of our study by simulating PTA data and performing parameter estimation for CGW targets and noise~\cite{vanHaasterenLevin2009}. 
PTA data is a set of radio pulsar pulse arrival times with the timing model, which provides an initial prediction of these arrival times~\cite{LuoRansom2021}. 
When fitting our models to the PTA data, we perform an analytical marginalisation over parameters of the timing model~\cite{ng9_gwb}.
We simulate future data of the Chinese Pulsar Timing Array (CPTA)~\cite{cpta_dr1_2023}.
We prioritise this over simulating data from long-established PTAs like the NANOGrav~\cite{ng15_timing}, the PPTA~\cite{ppta_dr3_timing}, the EPTA~\cite{epta_dr2_timing}, and their combined data as part of the International Pulsar Timing Array (IPTA)~\cite{ipta_dr2_data}, as it provides the opportunity to demonstrate the capabilities of our method with fewer simulated observations of pulsars.
Pulsar observations with the FAST telescope, utilised by the CPTA, achieve remarkable timing precision, thus enabling stringent constraints on CGW parameters with fewer observations, reducing the computational cost of the analyses. 

One of the objectives of our simulations is to perform an accurate ground-level representation of the expected levels of noise in future datasets. 
The white noise is simulated to match the root-mean-square residuals between the best-fit time series prediction of the timing and red noise models and the data for the first CPTA data release~\cite{ChenXu2025}. 
Pulsar-specific time-correlated ``red'' noise, which we model to exhibit a power law power spectral density of timing delays, is simulated to approximately represent its level of visibility and characteristics across pulsars, with a fiducial choice of the prior on its log-10 amplitude in units of GW strain, $\log_{10}A$, to be a truncated normal distribution $\mathcal{N}_\text{t}(-16.6, 1.8, -19.0, -10.0)$.
The spectral index, $\gamma$, follows $\mathcal{N}_\text{t}(2.5, 1.1, 1.0, 7.0)$, where the distribution parameters are the mean, standard deviation, minimum, and maximum, respectively.
Additionally, all-pulsar red noise simulates the effect of Fourier harmonics of the GWB with $\log_{10}A=-14.64$ and $\gamma=4.33$ based on EPTA observations~\cite{GoncharovSardana2025}.
The Hellings-Downs correlations of the GWB are neglected to reduce the computational burden. 
We do not simulate noise with properties depending on the radio frequency, such as stochastic fluctuations in pulsar dispersion measure.

\begin{figure}
    \centering
    \includegraphics[width=0.48\textwidth]{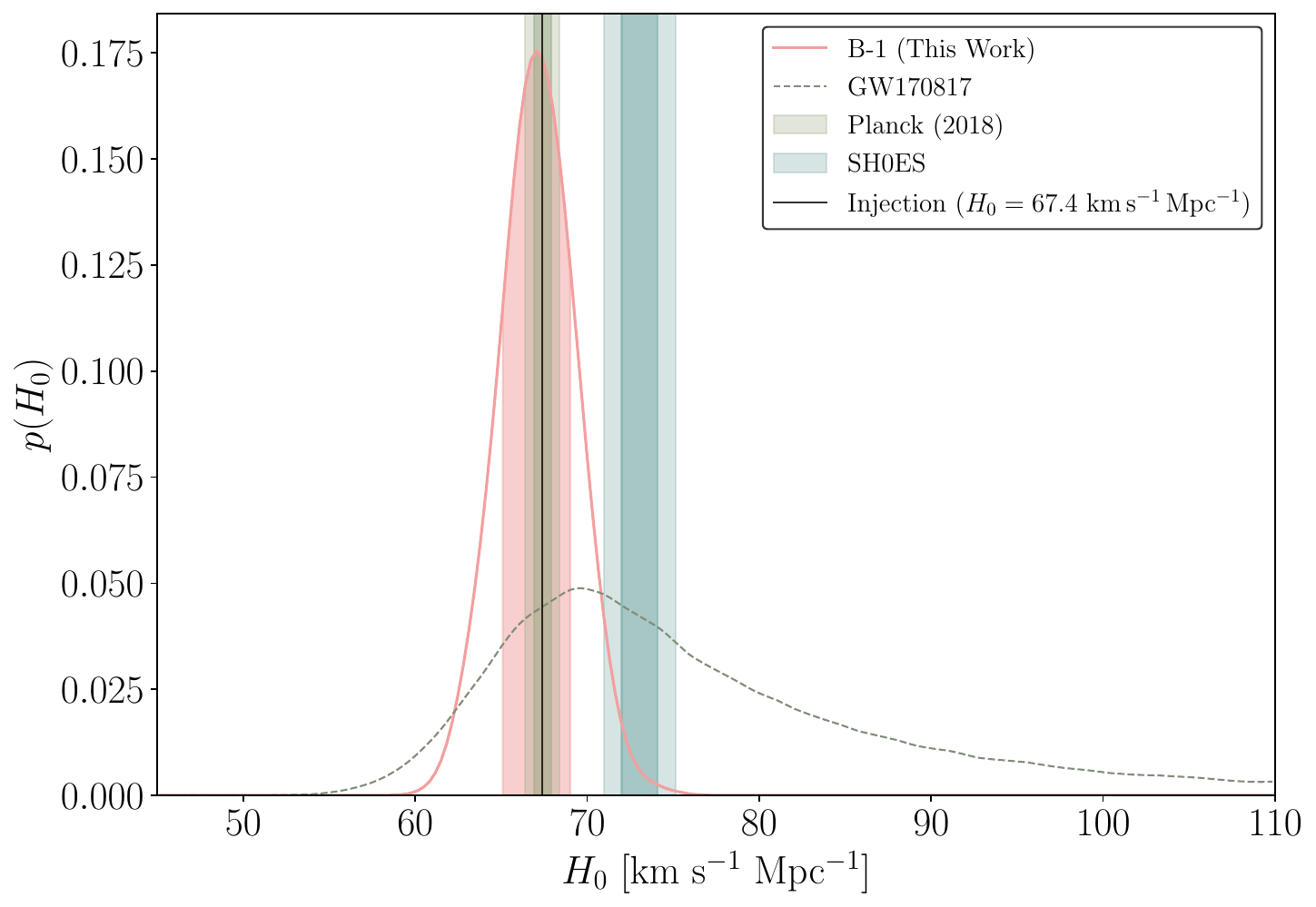}
    \caption{\justifying Marginalised posterior distributions for Hubble constant $H_0$. The result of our analysis with 40 best pulsars of the simulated 30-yr CPTA data (solid pink) is compared with the $H_0$ measurements from GW170817 (dashed grey), Planck 2018 (shaded olive), and SH0ES (shaded viridian). 
    The vertical black line indicates the simulated value $H_0 = 67.4 \text{ km s}^{-1} \text{ Mpc}^{-1}$.
    }
    \label{fig:H0_uncertainty}
\end{figure}

Demonstrating the remarkable capability of targeted standard sirens, we find $H_0=67.13^{+2.16}_{-2.16} \text{ km s}^{-1} \text{ Mpc}^{-1}$ in the analysis of the simulated $30$-year CPTA observation of a $3.69 \times 10^{9}~M_\odot$ SMBHB at SDSS~J072908.71+4008 (327.6~Mpc, z=0.07), referred to as B1 (SNR=95.642), with $40$ best\footnote{Pulsars are considered the best based on the white noise levels, although parameter estimation precision also strongly depends on pulsar locations with respect to the CGW source in the sky~\cite{RomanoAllen2024}.} pulsars. 
This result is achieved through reparametrising the CGW model~\cite{Ellis2013} such that the luminosity distance $D_\text{L}$ is replaced with the Hubble constant $H_0$ and the known redshift of B1's host galaxy. 
The simulated value of $H_0$ chosen to match the best-fit value from the Planck measurement of anisotropies in the cosmic microwave background~\cite{planck_2021}. 
The result of parameter estimation is shown in Figure~\ref{fig:H0_uncertainty}. 
The measurement precision of $H_0$ exceeds that of the first standard siren GW170818~\cite{gw170817_h0}, as well as the combined measurement with GW170818 and the VLBI~\cite{HotokezakaNakar2019}. 
In the subsequent subsections, we present more detailed calculations indicating the impact of the observation time, the number of pulsars, and the number of observed CGW sources on the measurement.

\subsection*{Cosmology with multiple simultaneously observed SMBHBs}

\begin{figure}[!htbp]
    \centering
    \begin{subfigure}[t]{0.48\textwidth}
        \centering
        \includegraphics[width=\linewidth]{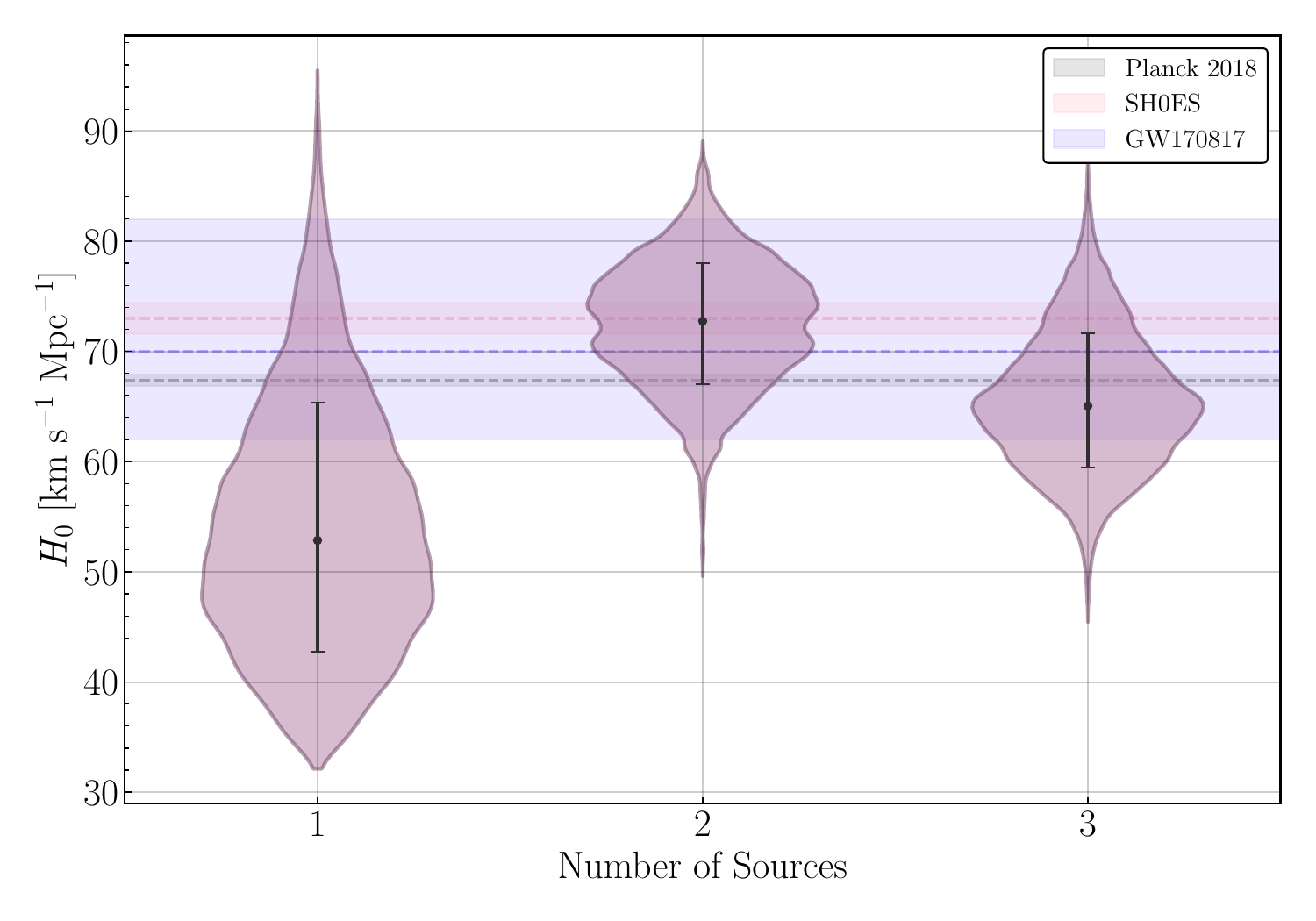}
    \end{subfigure}
    \caption{\justifying 
    Marginalised posteriors on $H_0$ for the simulated observation of SMBHB targets P1, P2, and P3. 
    The horizontal axis corresponds to the addition of these sources to the simulation one by one. 
    Central points and vertical error bars denote the median and $1\sigma$ credible intervals, respectively. Horizontal shaded regions indicate $1\sigma$ constraints from other cosmological probes: Planck 2018 (grey), SH0ES (pink), and the GW170817 standard siren measurement (purple).
    }
    \label{fig:sources_pulsars}
\end{figure}

We explore the full potential of the method by demonstrating the inference of $H_0$ with multiple detected SMBHB targets. 
Furthermore, to provide a perspective on the variability of our predictions concerning the measurement precision of $H_0$ with the CPTA, we simulate a set of other, quieter, SMBHB targets using population synthesis.  
We use the modified semi-analytical model ``Phenom$+$Uniform'' from Ref.~\cite{ng15pop}. 
Without modifications, this population yields a stochastic gravitational wave background with the strain amplitude of about $1.5\times10^{-15}$ at a frequency of $\text{yr}^{-1}$. 
To approximately match the strain amplitude of the stochastic gravitational wave background produced by this population model with the observed amplitude of $2.3\times10^{-15}$ at a frequency of $\text{yr}^{-1}$~\cite{GoncharovSardana2025}, we only make one modification. 
We change the log-10 normalisation of the galactic black hole at the reference bulge mass to be $8.6~[\log_{10}M_\odot]$ instead of the default value of $8.3~[\log_{10}M_\odot]$. 
Selecting the brightest sources across $60$ frequency bins with the initial frequency and the resolution of $20~\text{yr}^{-1}$, we identify the three most visible CGW sources: P1, P2, P3 (SNR=140.676, 4.392, 1.659 respectively for 20 best pulsars observed for 40 years). 
Optimistically, we assume that all three sources are valid SMBHB targets with known redshifts. 
Figure~\ref{fig:sources_pulsars} displays the marginalised posteriors on $H_0$ as we consequently add P1, P2, and P3 to the simulated data. 
We find that adding P2 drastically improves the precision. 
Adding a quieter source, P3, makes almost no visible difference. 
The dimensionality of our calculations rapidly increases with each additional injected CGW source, increasing the computational costs.

\section*{\label{sec:discussion} Discussion}

\subsection*{\label{sec:discussion:hostgal_misspec} Host Galaxy Misspecification}

For the targeted standard siren approach we introduced here, the risk of misidentification of the host galaxy requires particular attention.
In our analysis, we assumed a detection of SMBHB targets with the PTA data. 
However, the GW information alone does not allow a confirmation that the origin of the signal is a specific galaxy. 
A fiducial PTA sky localisation of about $10~\text{deg}^2$ for an all-sky blind search contains millions of galaxies. 
It is also well known that many of the SMBHB candidates found in electromagnetic observations may be false positives, meaning they do not emit the gravitational-wave signal with the properties expected by the targeted PTA searches. 
However, the scarcity of the SMBHBs at masses above $10^9~M_\odot$ and frequencies above $10^{-8}$~Hz, such as the signals we consider in this work, leads to the following conclusion. 
If such an SMBHB is detected in the targeted search, inferred parameters of this SMBHB are unlikely to simultaneously match parameters of any other SMBHB in the universe. 
Thus, either the SMBHB target and thus the host galaxy is valid, or the signal would otherwise become inconsistent with the parameters predicted by the EM observations. 

\begin{figure}[!ht]
    \centering
    \includegraphics[width=0.96\linewidth]{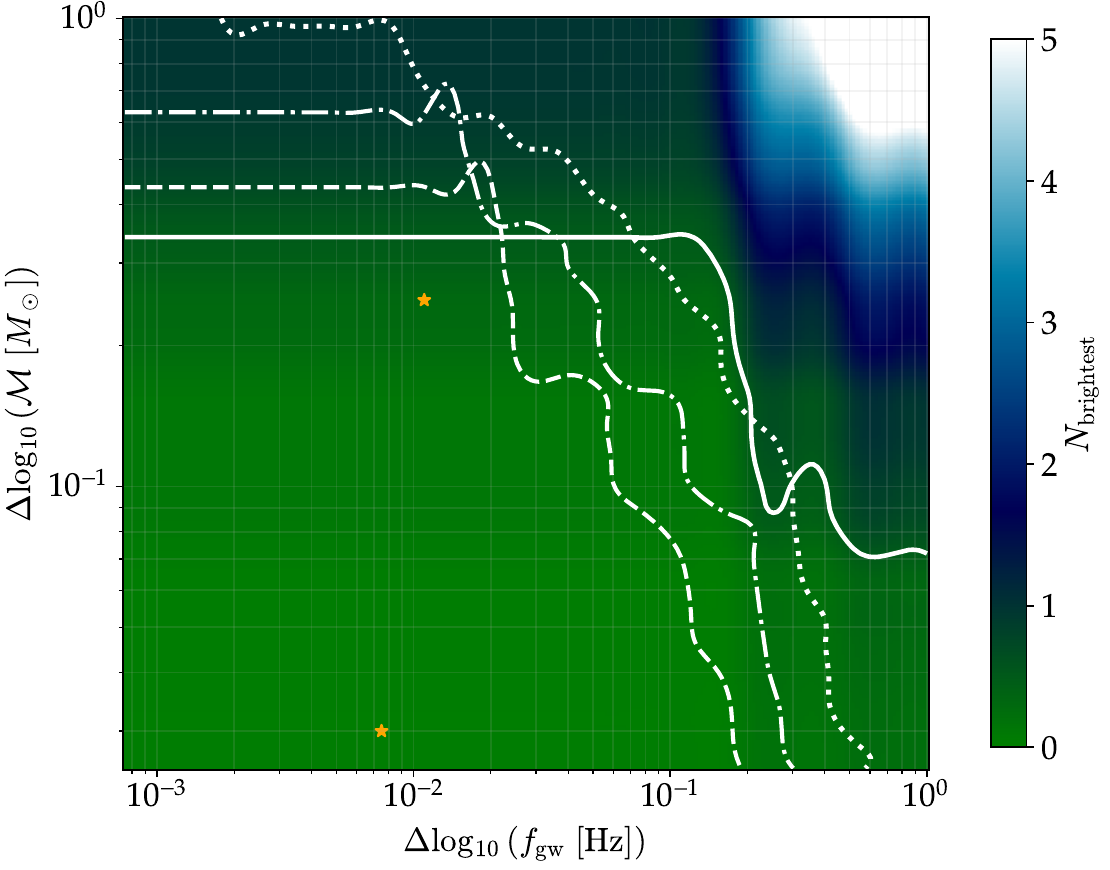}
    \caption{\justifying The average number of brightest CGW sources in the universe, $N_\text{brightest}$, shown as a function of assumed parameter estimation uncertainties ($\Delta$) with the PTA all-sky search for CGWs. 
    Values $N_\text{brightest}<1$ correspond to a low likelihood of confusion between the targeted SMBHB and any other plausible SMBHB, solely based on the match of their $\log_{10}f$ and $\log_{10}\mathcal{M}$. 
    The colour corresponds to a source P1. 
    The solid white line corresponds to a threshold for P1 where $N_\text{brightest}=0.5$. 
    The dashed, the dash-dotted, and the dotted lines correspond to the same threshold for P2, P3, and 3C~66B, respectively. 
    The top star and the bottom star correspond to reference parameter estimation uncertainties from~\cite{CharisiTaylor2024} and~\cite{PetrovSchult2025}, respectively. 
    The smallest value along the axes corresponds to $3\sigma$ credible levels in our simulated all-sky search for B1 with the $40$-year CPTA data. 
    }
    \label{fig:pop_uncertainties}
\end{figure}

To quantify the probability of SMBHB source misspecification, we simulate $100$ realisations of the brightest CGW sources based on the same semi-analytical model from which we obtained synthetic SMBHB sources (P1, P2, P3). 
Masses of SMBHBs in the simulated population span many orders of magnitude from $10^7$ to $10^{10}~M_\odot$, whereas redshifts are in the range between $0.01$ and $0.90$.
Let us assess the probability for our SMBHB candidates (P1, P2, P3) to be confused with other SMBHBs from this population. 
We propose that for the confusion to occur, the chirp mass and the frequency of SMBHBs from the population must all fall into the $2$-$3\sigma$ credible level of the posterior on target CGW parameters inferred from the all-sky search. 
In this case, we could still incorrectly identify the source of the signal as the CGW target, assuming an incorrect redshift of the target source. 
In Figure~\ref{fig:pop_uncertainties}, we show how many simulated sources from the population, averaged over $100$ realizations, are consistent with the inferred chirp mass and frequency of an SMBHB target, assuming a range of parameter estimation uncertainties for $\log_{10}\mathcal{M}$ and $\log_{10}f$. 
The smallest value along the axes approximately corresponds to the $3\sigma$ credible levels that we find for B1 in the all-sky search with the 40-year CPTA data with 40 best pulsars. 
The largest values along the axes correspond to an uncertainty of one order of magnitude in chirp mass and frequency. 
It is visible that, on average, there is much less than one source from the population that could be confused with B1. 
Thus, the probability of assuming an incorrect redshift for the targeted standard siren measurement of $H_0$ with B1 in the simulated $40$-year CPTA data is very low. 
For comparison, we show measurement uncertainties achieved in all-sky searches from other literature. 
Petrov~\textit{et al.}~\cite{PetrovSchult2025} report $1\sigma$ precision exceeding $\Delta(\log_{10}f,\log_{10}\mathcal{M})=(0.0075,0.0300)~\text{dex}$ in Figure~2 for the $20$-nHz $10^9~M_\odot$ simulated signal in the near-future 22-year IPTA data. 
Whereas Charisi~\textit{et al.}~\cite{CharisiTaylor2024} report a $2\sigma$ precision of about $(0.0110,0.2500)~\text{dex}$ for a slightly heavier SMBHB at a similar frequency in the synthetic 20-year NANOGrav data. 
The lines enclose areas with parameter estimation uncertainties required to obtain, on average, $<0.5$ sources from the population for P1, P2, P3, and 3C~66B, respectively. 
Measurement uncertainties achieved in Refs.~\cite{PetrovSchult2025} and~\cite{CharisiTaylor2024} are enclosed in these areas, highlighting that they are sufficient to avoid host galaxy misspecification in targeted CGW searches. 
Among the sources from the population that overlap with an SMBHB target in the parameter space, some sources may still be too distant to be detected. 
Other sources may have an inconsistent sky position. 
Thus, the estimates we outline in this paragraph are sufficiently conservative, and it may be possible to further revise the probability of source misspecification based on the PTA sensitivity. 
Furthermore, although the results presented in~\ref{fig:pop_uncertainties} are dependent on uncertainties in the population properties of SMBHBs, they provide a robust intuition for the validity of the approach we followed in this work.  
The first measurement of the Hubble constant with real PTA data would benefit from a more systematic study, as well as consistency tests between population properties of SMBHBs inferred from the gravitational wave background and individual sources. 

One may also consider mitigation techniques for the cases where parameter estimation is insufficient to assume that the SMBHB target is correctly specified. 
In these cases, the search boils down to the conventional PTA standard siren approach with all-sky CGW searches. 
Namely, we should consider that the observed CGW signal is from one of the galaxies within the sky position uncertainty, potentially filtering out galaxies that are unlikely to host SMBHBs with the chirp mass inferred from the PTA data. 
Alternatively, one may rely solely on the PTA data and multiple CGWs detected in targeted searches, introducing dropout parameters for SMBHBs, which yield parameter estimation for $H_0$ inconsistent with that provided by other CGWs.
This can be achieved with the use of dropout techniques and hierarchical models.

\subsection*{The impact of the timespan of the observation and the number of pulsars}

It is of interest to consider how the constraints on $H_0$ change as PTAs observe more pulsars for a longer duration. 
Longer observations improve the signal-to-noise ratio and the GW frequency resolution.
These improvements help identify the Earth and pulsar terms and better measure the chirp mass $\mathcal{M}$, which influences both CGW's amplitude and phase. 
Measuring $\mathcal{M}$ through the phase information allows to break the degeneracy between $\mathcal{M}$ and the luminosity distance $D_\text{L}$ in CGW's strain amplitude, $h_0 \propto \mathcal{M}^{5/3}/D_\text{L}$. 

\begin{figure*}[!htbp]
    \centering
    \begin{subfigure}[t]{0.48\textwidth}
        \centering
        \includegraphics[width=\linewidth]{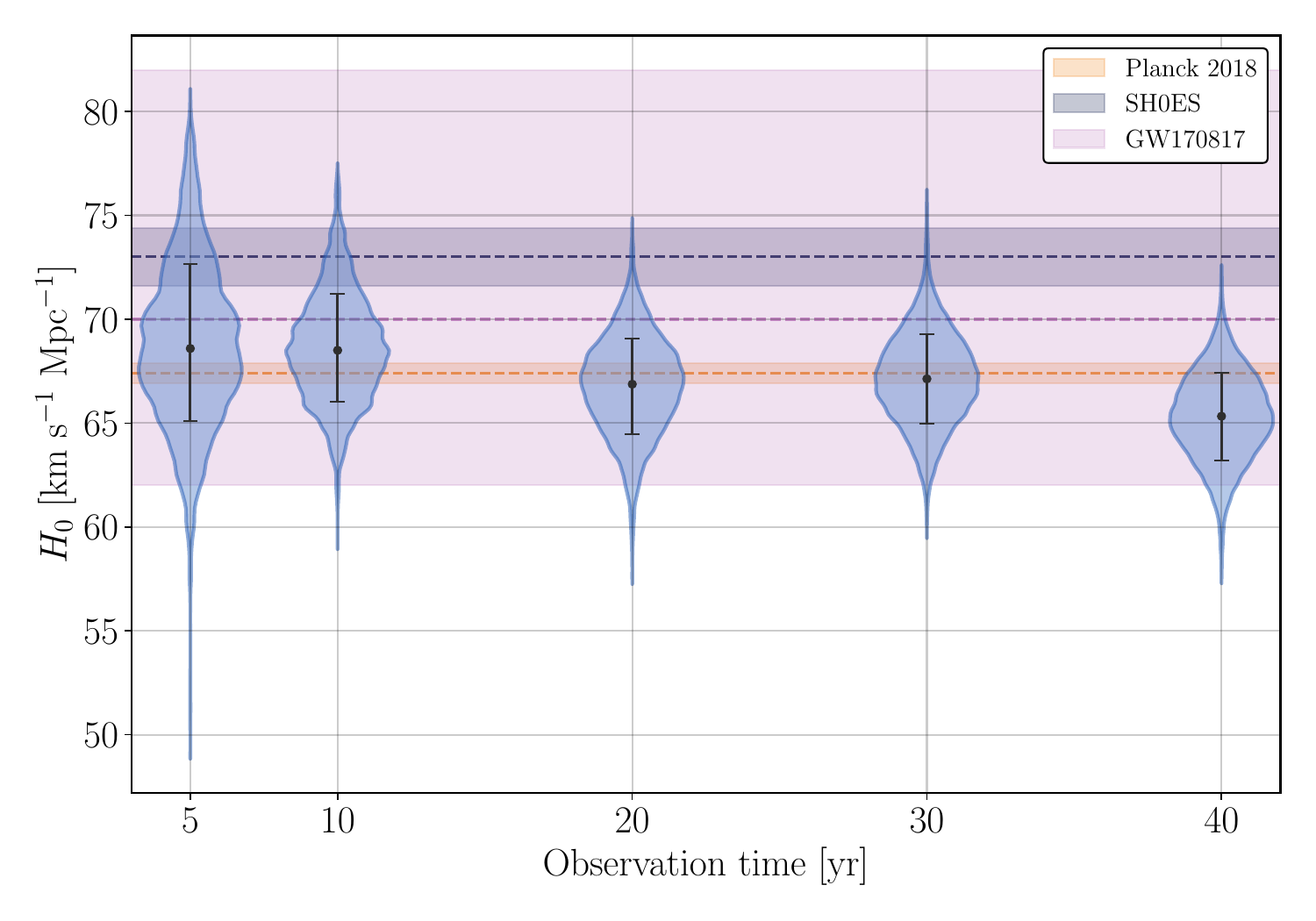}
        \label{fig:constraints_time}
    \end{subfigure}\hfill
    \begin{subfigure}[t]{0.48\textwidth}
        \centering
        \includegraphics[width=\linewidth]{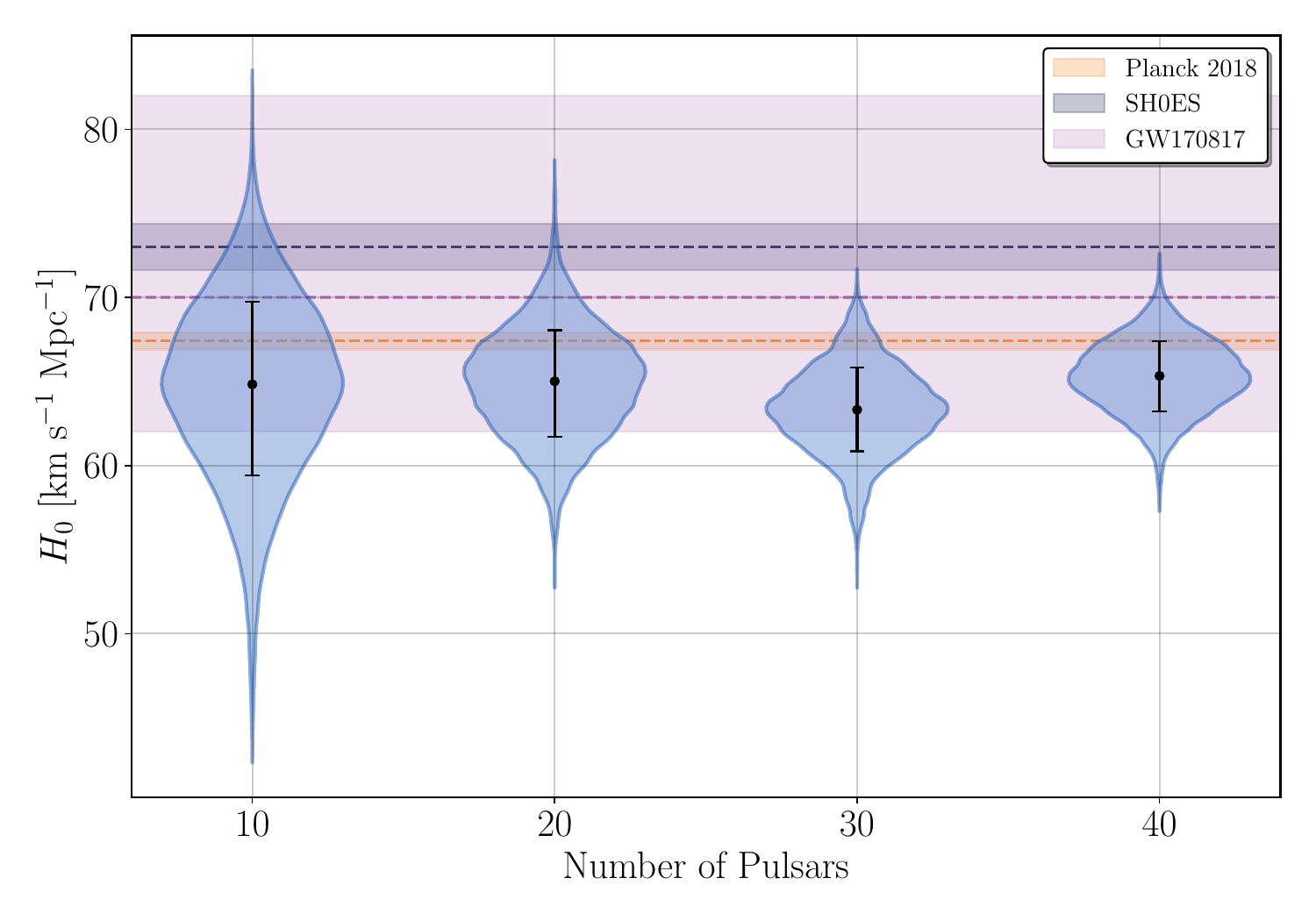}
    \end{subfigure}
    \caption{\justifying 
    The \textbf{left} panel shows the posterior for the Hubble constant $H_0$ as a function of observation time in the simulated targeted search for the SMBHB candidate B1 with 20 best pulsars of the CPTA. Central points and vertical error bars denote the median and $1\sigma$ credible intervals, respectively. 
    The \textbf{right} panel shows the marginalised posterior for $H_0$ for an increasing subset of pulsars used in the analysis. 
    Horizontal shaded regions indicate $1\sigma$ constraints from other cosmological probes: Planck 2018 (orange), SH0ES (purple), and the GW170817 standard siren measurement (pink).
    }
    \label{fig:time_pulsars}
\end{figure*}

Let us inspect posterior distributions of $H_0$ from simulated observations of a hypothetical SMBHB candidate B1, considering 5- to 40-year-long timing of the 20 best pulsars in the CPTA.
In Figure~\ref{fig:time_pulsars}, left, we show marginalised posteriors on $H_0$ as a function of the observation span. 
The precision of the $5$-year observation for 40 best CPTA pulsars exceeds that of the first observed standard siren, GW170817~\cite{gw170817_h0}, including the improved estimate based on the VLBI\footnote{Very-Long-Baseline interferometry.} observation of the jet in GW170817~\cite{HotokezakaNakar2019}. 
To clarify precisely how loud the CGW signal needs to be to achieve these constraints, we calculate an analytical all-sky signal-to-noise ratio (SNR) of B1~\cite{HazbounRomano2019}.

To investigate the influence of the number of pulsars in the CPTA on the precision of the inferred Hubble constant, we choose the 10 best pulsars based on their timing precision and increase the sample by 10 pulsars at a time.
We assume a $40$-year observation of the same CGW source as previously (B1). 
Figure~\ref{fig:time_pulsars}, right, displays the inferred $H_0$ and chirp mass $\mathcal{M}$ values with 1$\sigma$ credible levels for an increasing number of pulsars. 
While the addition of noisy pulsars does not contribute to the sensitivity of the array substantially, it improves the array's ability to break $\mathcal{M}$-$D_\text{L}$ degeneracy by increasing the number of orientations of pulsar-Earth baselines towards the CGW, improving the estimation of the Hubble parameter through better constraints on the luminosity distance.

\subsection*{\label{sec:discussion:pdist} Pulsar distances}

\begin{figure}
    \centering
    \includegraphics[width=0.48\textwidth]{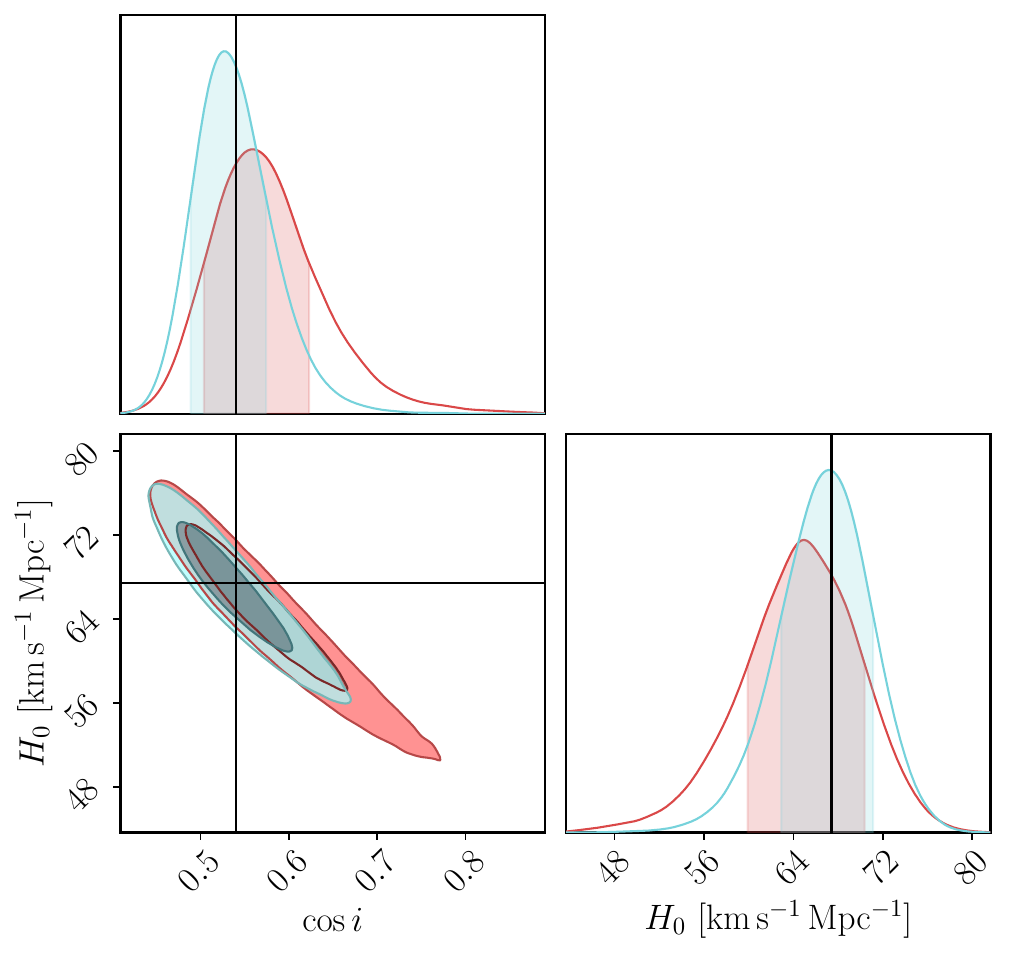}
    \caption{\justifying A demonstration of the impact of pulsar distance uncertainties on our parameter estimation for $H_0$ and the cosine of the inclination angle $\iota$ with the $40$-year observation of the 10 best CPTA pulsars. The posterior in red corresponds to our standard model, where pulsar distances are fitted. The posterior in blue corresponds to the case where pulsar distances are fixed.}
    \label{fig:pdist_zero}
\end{figure}

Our knowledge of pulsar distances determines how well we measure $\mathcal{M}$, breaking the degeneracy between $\mathcal{M}$ and $D_\text{L}$. 
Let us determine to what extent it is true. 
In this work, for most of the pulsars, we assume pulsar distances to be 1000~pc with 20\% uncertainty, corresponding to typical values realised in practice. 
The uncertainties on pulsar distances, modelled as Gaussian standard deviations, are fitted parameters. 
As a toy example, let us consider that we know pulsar distances with absolute certainty in the inference of $H_0$ with 40 years of observations of the 10 best pulsars of the CPTA. 
The result of parameter estimation of $H_0$ and a cosine of the inclination angle $\iota$ is shown in Figure~\ref{fig:pdist_zero}. 
It is compared against the result obtained with the standard assumption about pulsar distances. 
The posterior on $\mathcal{M}$ is not shown because the uncertainty on this parameter reduces very drastically, such that it becomes invisible compared to the posterior obtained with fitted pulsar distances. 
A degeneracy between $\mathcal{M}$ and $D_\text{L}$ in this simulation is also already mostly resolved. 
As a result, we obtain about $22$\% improvement in the uncertainty on $H_0$. 
This demonstrates that improving pulsar distance measurements is still very helpful. 
For less informative observations, a precise knowledge of pulsar distances would introduce a greater improvement. 
A posterior with 40 years of observations of the 10 best pulsars of the CPTA, highlighting the lack of strong degeneracy between $\mathcal{M}$ and $D_\text{L}$, is shown in the Appendix. 

\subsection*{\label{sec:discussion:pphase} CGW phases at pulsars}

\begin{figure}
    \centering
    \includegraphics[width=0.48\textwidth]{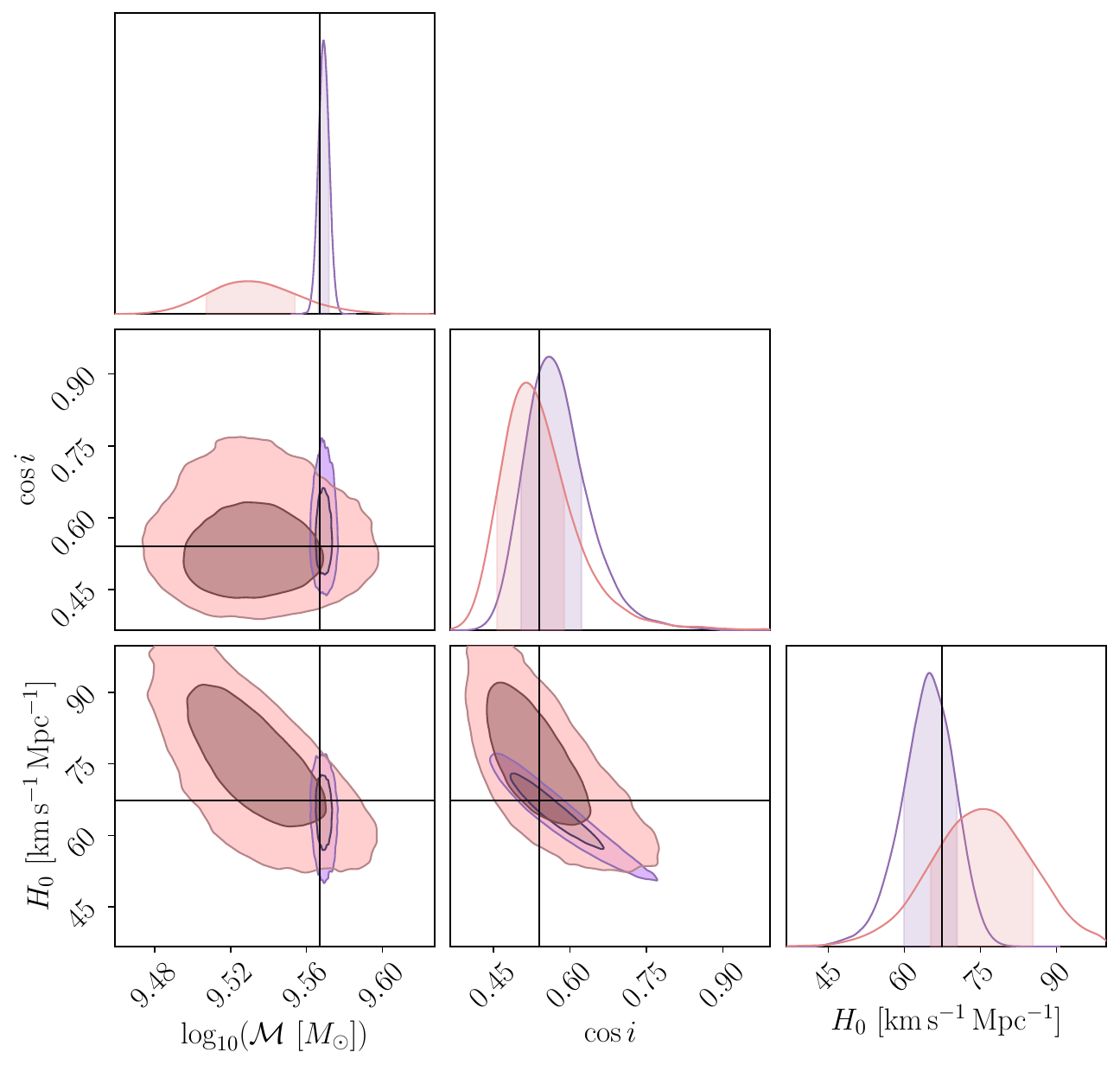}
    \caption{\justifying An impact of imposing a redundant CGW pulsar phase uncertainty. Posterior comparison between the cases where the CGW phase at the pulsar is fit (pink) and where it is not (purple).}
    \label{fig:normal_v_pphase}
\end{figure}

Let us now assess the role of the CGW phase at a pulsar. 
The CGW signal time series at pulsar position is uniquely determined by our knowledge of the pulsar's sky position, distance, and frequency evolution (``chirping'') due to the binary orbital energy loss. 
With the redundant pulsar-specific phase parameters, we introduce an additional uncertainty of up to one full CGW cycle at pulsars. 
The result of parameter estimation with pulsar phase parameters is compared against our standard parameter estimation in Figure~\ref{fig:normal_v_pphase}. 
Marginalisation over an imposed uncertainty in the CGW pulsar phase results in a significant reduction in the precision of the estimated $\mathcal{M}$, and to a lesser reduction in the precision of the estimated $H_0$. 
As in the case of the pulsar distance estimation, the effect depends on the level of degeneracy between $\mathcal{M}$ and $D_\text{L}$. 
Provided that the distances to pulsars are on the order of hundreds of light-years, whereas the CGW signal wavelengths are on the order of light-years, CGW signal models need to accurately predict the phase at the pulsar with high precision over hundreds of cycles.

\section*{Conclusion}

Supermassive binary black hole (SMBHB) candidates selected in active galactic nuclei (AGN) observations can serve as standard sirens to infer $H_0$ if their nature is confirmed through the detection of continuous gravitational waves (CGWs) with Pulsar Timing Arrays (PTAs).
In this work, we consider observing scenarios where between one and three such candidates are confirmed as genuine SMBHBs. 
Although many SMBHB candidates may be false positives, they are more likely to contain SMBHB than random galaxies~\cite{Casey-ClydeMingarelli2024}. 
Therefore, a significant number of SMBHB candidates may be confirmed. 

The CGW source with the most precise luminosity distance measurement is likely to provide a dominant contribution to the measurement of $H_0$. 
Our simulations show that achieving a precision of $2~\text{km}~\text{s}^{-1}~\text{Mpc}^{-1}$ is possible with the confirmed SMBHB target with a chirp mass $\mathcal{M}=3.69 \times 10^9~M_\odot$ at $z=0.07$ with the CPTA alone. 
These parameters, as well as the sky position we use in the simulation, are based on the SMBHB candidate B1 (SDSS~J072908.71+4008).
B1 is one of $114$ SMBHB candidates followed up by NANOGrav, for which they find a Bayes factor of 3. 
In addition to the presence of our simulated source in the AGN SMBHB catalogue, we find that B1 is consistent with the population synthesis model we have used to simulate SMBHB targets P1, P2, and P3~\cite{ng15pop}. 

We have systematically explored the limiting factors in our analysis, specifically the number of observed pulsars, the population of CGW sources, the total observation timespan, and the uncertainty in pulsar distances. 
We find that intrinsic pulsar red noise, sampling cadence, and TOA errors remain as the dominant factors limiting the accumulation of the signal-to-noise ratio (SNR). 
The most significant bottleneck in improving the measurement precision of $H_0$ is the TOA precision. 
Without achieving sub-100-ns accuracy, PTA sensitivity remains insufficient to treat SMBHBs as standard sirens. 

Whether the precision of $2~\text{km}~\text{s}^{-1}~\text{Mpc}^{-1}$ is going to be achieved or even exceeded depends on a range of other factors. 
The most important factor is the abundance of nearby SMBHBs, which stand out from the gravitational wave background, as well as the fraction of these bright SMBHBs as candidates determined based on AGN observations. 
While this is not discussed in the results section, we achieve the precision of $1~\text{km}~\text{s}^{-1}~\text{Mpc}^{-1}$ when supplementing a 40-year 40-pulsar CPTA observation of B1 with an observation of another source, B2 (another SMBHB outline in the NANOGrav targeted search associated with SDSS J153636.22+044127.0). 
However, three promising sources we found in the first stochastic realisation of our population synthesis simulation only yield a precision of $5~\text{km}~\text{s}^{-1}~\text{Mpc}^{-1}$. 
Moreover, observations of pulsars across a wide range of radio frequencies and accurate models of the interstellar propagation effects are necessary to ensure our predictions obtained without considering these effects are close to reality.  
With this, we have conservatively modelled the gravitational wave background as a power-law, whereas a realistic background is expected to decline at most of our CGW frequencies while forming CGWs. 
This effect would reduce the noise level that the background creates in our measurement of $H_0$. 
Additionally, we have only considered observations with the CPTA, neglecting the possibility of data combinations with the existing and future PTAs, which will utilise other state-of-the-art instruments like DSA-2000~\cite{HallinanRavi2019} and the Square Kilometre Array (SKA)~\cite{ShannonBhat2025}. 
Resolving a distance-inclination degeneracy which persists even with $40$-year observations can be potentially mitigated with VLBI astrometry of the jet emitted by a component of the SMBHB.

\section*{\label{sec:methods} Methodology}

\subsection*{Cosmological distance measurements with PTAs}

We assume a standard flat $\Lambda$CDM cosmology primarily governed by cold dark matter (CDM) with a cosmological constant ($\Lambda$), where the expansion history of the Universe is described by the Hubble parameter $H(z)$:
\begin{equation}
    H(z) = H_0 \sqrt{\Omega_\text{m} (1+z)^3 + \Omega_{\Lambda}},
\end{equation}
where $H_0$ is the Hubble constant~\cite{Peebles_book1994}, $\Omega_\text{m}$ and $\Omega_\Lambda$ denote the present-day dimensionless density parameters for matter and dark energy, respectively. For a flat universe, these parameters satisfy $\Omega_\text{m} + \Omega_\Lambda = 1$. In this work, we adopt the Planck 2018 results~\cite{planck_2021}, fixing $\Omega_\text{m} = 0.315$ and $\Omega_{\Lambda} = 0.685$.\\
The primary observable bridging a CGW signal to cosmology is the luminosity distance $D_{\text{L}}$. 
For a potential CGW source in the local universe ($z \ll 1$), $D_{\text{L}}$ can be linearly approximated via the Hubble law:
\begin{equation}
    D_{\text{L}} \approx \frac{cz}{H_0}.
    \label{eq:approx_hubble_equation}
\end{equation}
However, since PTA sources can span a wide range of redshifts ($z \lesssim 2.5$), we must use the full integral expression for a flat $\Lambda$CDM universe~\cite{Hogg1999}:
\begin{equation}
    D_{\text{L}}(z, H_0) = \frac{c(1+z)}{H_0} \int_0^{z} \frac{dz'}{\sqrt{\Omega_\text{m}(1+z')^3 + \Omega_{\Lambda}}}.
    \label{eq:lcdm_dist}
\end{equation}
Redshift measurements in electromagnetic galaxy catalogues ($z_{\text{obs}}$) include contributions from both the cosmological expansion (Hubble Flow, $z_{\text{flow}}$) and the host galaxy’s peculiar velocity ($v_{\text{p}}$). These are related by~\cite{DavisScrimgeour2014}:
\begin{equation}
    (1+z_\text{obs}) = (1+z_\text{flow})\left(1 + \frac{v_\text{p}}{c}\right).
\end{equation}
The contribution of peculiar velocity to the observed redshift is more significant relative to the Hubble flow for sources in the nearby universe than for those at high redshifts~\cite{HutererShafer2017}. 
Consequently, accounting for $v_\text{p}$ becomes increasingly important at lower $D_{\text{L}}$. Furthermore, as distance increases, the independent determination of $v_\text{p}$ relies on distance indicators that become increasingly uncertain with the distance. This creates a systematic floor in the error budget, propagating uncertainties into both $D_{\text{L}}$ and $H_0$ estimations. 
In this study, we sidestep these complexities by utilising simulated datasets where CGW sources strictly follow the Hubble flow ($v_{\text{p}} = 0$).

\subsection*{Redshift effects on SMBHB parameters }
The binary properties observed by PTAs, such as chirp mass ($\mathcal{M}_\text{obs}$) and the inspiral frequency ($f_\text{obs}$), are frame-dependent quantities. The observed chirp mass ($\mathcal{M}_\text{obs}$) is related to the rest-frame (source) Chirp mass at a redshift $z$ by~\cite{epta_dr2_cw}
\begin{equation}
    \mathcal{M}_\text{obs} = (1+z)\mathcal{M}_\text{src}.
\end{equation}
Similarly, the frequency of binary inspiral in the observer's frame ($f_\text{obs}$) is written as
\begin{equation}
    f_\text{obs} = \frac{f_\text{src}}{(1+z)}.
\end{equation}

For sources at high redshifts, SMBHBs appear more massive and evolve more slowly in the observer frame than in the source rest-frame. Since the gravitational-wave strain amplitude $h_0$ scales with the observed chirp mass as $h_0 \propto \mathcal{M}_{\text{obs}}^{5/3}$, redshift effectively boosts the signal strength for distant sources. 

\subsection*{Continuous Wave Signal Model}
Pulsar timing delays induced by a quasi-monochromatic, continuous signal from a circular and non-precessing SMBHB are expressed as a combination of the Earth term (at the observatory) and the Pulsar term (at the pulsar location)~\cite{Ellis2013}. The total timing delay $s(t, \hat{\Omega})$ induced by the CGW for a pulsar in the direction given by the unit vector $\hat{p}$ is
\begin{align}
    s(t, \hat{\Omega}) =
    F^{+}(\hat{\Omega}) \Delta s_{+}(t) + F^{\times}(\hat{\Omega}) \Delta s_{\times}(t),
    \label{eq:cgw_residuals}
\end{align}
where $\Delta s_{+, \times}(t) = s_{+, \times}(t_\text{p}) - s_{+, \times}(t)$, and $t$ and $t_\text{p}$ denote the times at which the CGW wavefront passes the solar-system-barycentre (SSB) and the pulsar, respectively. From geometric considerations, the time $t_\text{p}$ could be related to $t$ by
\begin{equation}
    t_\text{p} = t - \frac{L}{c}(1 - \cos(\mu)).
\end{equation}
where $L$ is the distance to the pulsar and $\mu$ is the angle between the GW propagation direction ($-\hat{\Omega}$) and the Earth-pulsar unit vector ($\hat{p}$). The term $(1 - \cos\mu)$ represents the geometric delay factor. For a CGW source at an inclination of $\iota$, the functions $s_{+,\times}$ describe the phase evolution of the wave polarisations and are expressed as
\begin{align}
    s_{+}(t) =\, 
    \frac{(G\mathcal{M})^{5/3}}{D_{\text{L}}c^4\omega(t) ^ {1/3}} 
    &[-\sin\!\left(2\phi(t)\right)\left(3+\cos^2\iota\right)\cos(2\psi)\notag\\
    &-\,2\cos\!\left(2\phi(t)\right)\cos\iota\,\sin(2\psi)],
\end{align}
\begin{align}
    s_{\times}(t) = \,
    \frac{(G\mathcal{M})^{5/3}}{D_{\text{L}}c^4 \omega(t) ^ {1/3}}
    &[-\sin\!\left(2\phi(t)\right)\left(1+\cos^2\iota\right)\sin(2\psi)\notag\\
    &+\,2\cos\!\left(2\phi(t)\right)\cos\iota\,\cos(2\psi)],
\end{align}

The orbital phase evolution $\phi(t)$ and frequency $\omega(t)$ are given by 

\begin{equation}
    \phi(t) = \phi_0
    + \frac{1}{32} \left (\frac{G \mathcal{M}}{c^3} \right )^{-5/3}
    \left(
    \omega_0^{-5/3} - \omega(t)^{-5/3}
    \right),
\end{equation}
and
\begin{equation}
    \omega(t) =
    \left(
    \omega_0^{-8/3}
    - \frac{256}{5}\, \left ( \frac{G \mathcal{M}}{c^3} \right )^{5/3} t
    \right)^{-3/8},
\end{equation}
where $\phi_0$ represents the initial phase and $\omega_0$ the initial orbital frequency~\cite{EllisSiemens2012,CorbinCornish2010}.   

The functions $F^{+}(\hat{\Omega}, \hat{p})$ and $F^{\times}(\hat{\Omega}, \hat{p})$ are the antenna response functions, describing the geometric sensitivity of the pulsar to the two GW polarizations from the CGW source at a sky location $\hat{\Omega}$. $F^{+}(\hat{\Omega}, \hat{p})$ and $F^{\times}(\hat{\Omega}, \hat{p})$ are defined as
\begin{equation}
    F^{+}(\hat{\Omega}) = \frac{1}{2} \frac{(\hat{m} \cdot \hat{p})^{2} - (\hat{n} \cdot \hat{p})^{2}}{1 + \hat{\Omega} \cdot \hat{p}},
\end{equation}
\begin{equation}
    F^{\times}(\hat{\Omega}) = \frac{(\hat{m} \cdot \hat{p})(\hat{n} \cdot \hat{p})}{1 + \hat{\Omega} \cdot \hat{p}},
\end{equation}
The vectors $\hat{m}$ and $\hat{n}$ are the unit vectors defining the principal axes of the GW polarization, both of which are orthogonal to the propagation direction ($-\hat{\Omega}$) and to each other as well. 

In this work, we explicitly model both the Earth and pulsar terms using Equation~\ref{eq:cgw_residuals}, thereby leveraging the phase and frequency evolution across a baseline of thousands of light-years. Incorporating the pulsar term into the waveform is essential for breaking the degeneracy between the chirp mass ($\mathcal{M}_\text{obs}$) and the luminosity distance ($D_{\text{L}}$). Pulsar term provides an additional measurement of the CGW phase at a much earlier epoch, introducing an extra constraint that breaks this degeneracy. This improves the sky localisation of the source and effectively enables a precise ``triangulation'' of the CGW source~\cite{CorbinCornish2010}.

\subsection*{Targeted CGW Cosmology}

A typical all-sky search is a blind search across the parameter space. It is based on finding the luminosity distance ($D_{\text{L}}$) of the CGW model as well as the chirp mass ($\mathcal{M}$), sky location ($\hat{\Omega}$), CGW frequency ($f_{\text{gw}}$), binary inclination with respect to the line of sight ($\iota$), and the initial phase ($\phi_0$) at a reference epoch $t_{\text{ref}}$. 

In contrast, in standard sirens based on targeted CGW searches, the CGW source primarily serves as a cosmic distance indicator for $H_0$, and the priors on other parameters are informed by electromagnetic (EM) observations of AGN. 
The EM-informed CGW parameters include the chirp mass $\mathcal{M}$, the sky position, and the gravitational wave frequency $f_\text{gw}$, and the CGW parameter $D_\text{L}$ is substituted with $(H_0,z)$. 
For B1, we assume $\pi(\log_{10}\mathcal{M})=\mathcal{N}(9.567,0.3)$ and $f_\text{gw}=14.4\times10^{-9}$, and $\pi(\log_{10}\mathcal{M})=\mathcal{N}(9.829,0.34)$ with $f_\text{gw}=20.8\times10^{-9}$ for B2. 
For the sources P1, P2, and P3 from the population analysis we use $\pi(\log_{10}\mathcal{M})=\mathcal{N}(10.019,0.3)$ and $f_\text{gw}=4.753\times10^{-9}$ for P1, $\pi(\log_{10}\mathcal{M})=\mathcal{N}(8.976,0.3)$ and $f_\text{gw}=2.852\times10^{-8}$ for P2, and $\pi(\log_{10}\mathcal{M})=\mathcal{N}(9.073,0.3)$ and $f_\text{gw}=3.803\times10^{-8}$ for P3.
The sky positions $\hat{\Omega}$ and redshifts $z$ are known to such a high precision that PTAs may not be able to exceed. The measurement precision for the CGW frequency is also high (see Figure~\ref{fig:pop_uncertainties} and Ref.~\cite{PetrovSchult2025}), so the frequency parameters are also fixed at simulated values during the parameter estimation, for simplicity.
The value of $H_0$ parameter is used to compute the luminosity distance $D_{\text{L}}(z, H_0)$ using $\Lambda$CDM (Equration~\ref{eq:lcdm_dist}). 
As a result, in targeted CGW searches, compared to all-sky CGW searches, we reduce the dimensionality of the parameter space and break the degeneracies that typically plague all-sky searches.

\subsection*{Simulated Data}

To validate our method, we simulate a measurement of the Hubble constant $H_0$ with the future Chinese Pulsar Timing Array (CPTA) by reporting constraints for different observation spans, considering a various number of simultaneously observed CGW sources, and different subsets of pulsars. 
For each pulsar and each synthetic data, we produce a random time series realisation of stochastic noise processes.
White noise is represented by the error factor parameter, EFAC$=1$, meaning that the delays induced by white noise are well described by the Gaussian distribution with the standard deviation matching measurement uncertainties in pulse arrival times. 
Red noise, both pulsar-intrinsic (RN) and all-pulsar (CURN), is modelled to exhibit a power law power spectrum. 
Additionally, we simulate continuous gravitational wave signals mimicking the specific CGW candidates. 
The selection of simulated CGW sources is motivated by the recent observation results by the NANOGrav 15-year dataset, which has shown minor evidence for the CGW signal from SDSS J072908.71+4008 (Gondor) and SDSS J153636.22+0441 (Rohan) binary candidates~\cite{ng15_cw_target}. 
\begin{figure}
    \centering
    \includegraphics[width=0.48\textwidth]{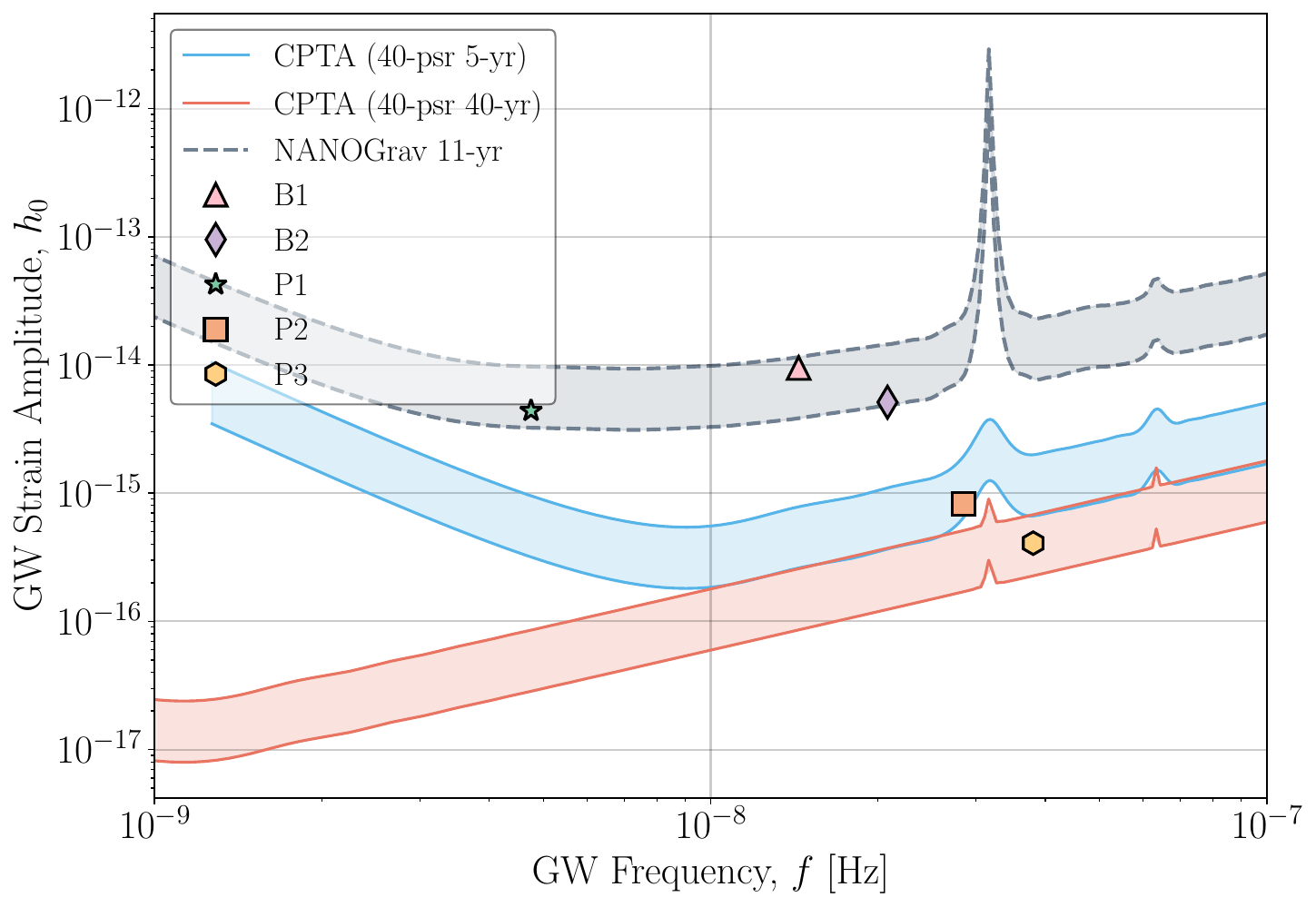}
    \caption{\justifying A plot comparing the strain amplitude of SMBHBs used in this analysis and the strain curve for 40 CPTA pulsars for 5 and 40 years of data. The shaded region represents the region between SNR=1 and SNR=3.}
    \label{fig:hasasia_plot}
\end{figure}
The simulated binary parameters and the noise properties of the pulsars in the CPTA are detailed in Table~\ref{tab:injection}.

\begin{table*}[!ht]
    \centering
    \renewcommand{\arraystretch}{1.5}
    \setlength{\tabcolsep}{17pt}
    \begin{tabular}{|l|l|l|}
    \hline
    \textbf{Source: ``P1''} ($z=0.3483$) & \textbf{Source: ``P2''} ($z=0.1266$) & \textbf{Source: ``P3''} ($z=0.3905$) \\
    \hline
    RA $= -05^{\text{h}}00^{\text{m}}10.0^{\text{s}}$ & RA $= 04^{\text{h}}03^{\text{m}}21.0^{\text{s}}$ & RA $= -00^{\text{h}}08^{\text{m}}09.0^{\text{s}}$ \\
    Dec $= +15^\circ01'12.0''$ & Dec $= +20^\circ32'7.0''$ & Dec $= +59^\circ05'18.0''$ \\
    $\mathcal{M}_c = 1.045\times10^{10}\,M_\odot$ & $\mathcal{M}_c = 9.458\times10^8\,M_\odot$ & $\mathcal{M}_c = 1.183\times10^{9}\,M_\odot$ \\
    $f_\text{gw} = 4.753$ nHz & $f_\text{gw} = 28.519$ nHz & $f_\text{gw} = 38.026$ nHz \\
    $D_\text{L} = 1910.495$ Mpc & $D_\text{L} = 615.040$ Mpc & $D_\text{L} = 2183.835$ Mpc \\
    $\iota = 57.316^\circ$  & $\iota = 45.0^\circ$  & $\iota = 52.0^\circ$  \\
    $\psi = 1.54$  & $\psi = 1.59$  & $\psi = 1.59$   \\
    $\phi_0 = 3.14$  & $\phi_0 = 3.17$  & $\phi_0 = 3.17$  \\
    \hline\hline
    
    \textbf{Source: ``B1''} ($z=0.07$) & \textbf{Source: ``B2''} ($z=0.38$) & \textbf{Noise Component} \\
    \hline
    RA $= -07^{\text{h}}29^{\text{m}}08.6^{\text{s}}$ & RA $= 15^{\text{h}}36^{\text{m}}36.2^{\text{s}}$ & White Noise: EFAC $= 1.0$ \\
    Dec $= +40^\circ08'37.0''$ & Dec $= +04^\circ41'26.9''$ & $\log_{10} A_\text{RN}: \mathcal{N}_\text{t}(-16.6, 1.8, -19.0, -10.0)$ \\
    $\mathcal{M}_c = 3.69\times10^9\,M_\odot$ & $\mathcal{M}_c = 6.745\times10^9\,M_\odot$ & $\gamma_\text{RN}: \mathcal{N}_\text{t}(2.5, 1.1, 1.0, 7.0)$ \\
    $f_\text{gw} = 14.4$ nHz & $f_\text{gw} = 20.8$ nHz & $\log_{10} A_\text{CURN} = -14.64$ \\
    $D_\text{L} = 327.575$ Mpc & $D_\text{L} = 2115.110$ Mpc & $\gamma_\text{CURN} = 4.33$ \\
    $\iota = 57.316^\circ$  & $\iota = 110.487^\circ$  & \\
    $\psi = 1.54$  & $\psi = 1.59$  & \\
    $\phi_0 = 3.14$  & $\phi_0 = 3.17$  & \\
    \hline
    
    \end{tabular}
    \caption{\justifying The summary of our simulated signals and noise. The first five panels list the physical parameters for the CGW sources (P1, P2, P3, as well as B1 and B2). 
    The final panel details the noise properties of the simulation.}
    \label{tab:injection}
\end{table*}
\subsection*{Likelihood}

We assume the pulsar pulse times of arrival (TOA) to be 
\begin{equation}
    \delta \mathbf{t} = \mathbf{M}\boldsymbol{\epsilon} + \mathbf{F}\mathbf{a} + \mathbf{s} + \mathbf{n}.
\end{equation}
Stochastic signals time series $\mathbf{F}\mathbf{a}$ and timing model contributions to TOAs $\mathbf{M}\boldsymbol{\epsilon}$ are written in a form of reduced-rank modelling~\cite{LentatiAlexander2013}, such that design matrices $(\mathbf{M},\mathbf{F})$ map vector coefficients $(\mathbf{a},\boldsymbol{\epsilon})$, representing Fourier power spectrum amplitudes and linear corrections to the timing model parameters, respectively, to time series. 
Vector $\mathbf{s}$ is the deterministic CGW waveform defined in Equation~\ref{eq:cgw_residuals}. 
Finally, $\mathbf{n}$ is the white noise time series.

The likelihood of the data is a multivariate normal distribution~\cite{vanHaasterenLevin2009} marginalised over coefficients $(\mathbf{a},\boldsymbol{\epsilon})$~\cite{ng9_gwb},
\begin{align}
    \ln \mathcal{L}
    &= -\frac{1}{2}
    \Big[
    (\delta \mathbf{t} - \mathbf{s})^{\mathrm{T}}
    \mathbf{C}^{-1}
    (\delta \mathbf{t} - \mathbf{s})
    \nonumber \\
    &\quad + \ln \det \left( 2\pi \mathbf{C} \right)
    \Big].
\end{align}
The covariance matrix, $\mathbf{C}$, contains contributions of stochastic noise processes, uncorrelated in time (white) and time-correlated (red) noise, to the TOA time series. 
In particular, 
\begin{equation}
    \mathbf{C} = \mathbf{N} + \mathbf{F}\mathbf{B}\mathbf{F}^\text{T}.
\end{equation}
Here, $\mathbf{N}$ is the diagonal covariance matrix (white noise), with elements $\sigma^{2}=\text{EFAC}^{2} \sigma^{2}_\text{TOA}$, where $\text{TOA}$ is a measurement uncertainty of TOAs simulated based on Ref.~\cite{ChenXu2025}. 
Matrix $\mathbf{B}$ is the reduced-rank covariance matrix containing the power spectral density amplitudes for both the intrinsic red noise and the all-pulsar common uncorrelated red noise (CURN) processes. 
To maintain computational efficiency, we utilise the Sherman-Morrison-Woodbury identity to evaluate both the inverse and the determinant of the covariance matrix $\mathbf{C}$.
Therefore, by performing matrix operations in the rank-reduced frequency basis, we avoid the high computational cost of directly inverting the large covariance matrix.

\subsection{Parameter whitening}

To improve sampler efficiency in high-dimensional parameter spaces, all the parameters are mapped to a unit Gaussian prior. For parameters with uniform or log-uniform priors, we applied the probit transform, $u = \Phi^{-1} \left((x-x_\text{min})/(x_\text{max}-x_\text{min}\right)$, where $\Phi$ is a standard normal cumulative density function (CDF). 
Parameters with Gaussian priors are mapped as $u = (x-\mu)/\sigma$, and parameters with truncated Normal priors are mapped as $u = \Phi^{-1}\left((\Phi((x-\mu)/\sigma) - \Phi_{\text{lo}})/\Delta\Phi\right)$ using conditional CDF mapping. 
Here, $\Delta\Phi = \Phi_{\text{hi}}-\Phi_{\text{lo}}$, where $\Phi_{\text{lo}}$ and $\Phi_{\text{hi}}$ are the CDF evaluated at the maximum and minimum truncation bound. 
Under this reparameterisation, all the parameters are passed as $\mathcal{N}(0,1)$ to the sampler, eliminating differences in prior scale across dimensions. 
The sampler is then run entirely in the whitened space, and the likelihood is evaluated by transforming samples back to the physical parameter space at each step. 
This adaptation removes the cold-start problem for the Adaptive Metropolis jump proposals and prevents ``overshooting'' in Differential Evolution jumps.

\section*{Data and code availability}

The data and the code generated as part of this work are available upon request from the corresponding author. The targeted search for continuous gravitational wave and parameter estimation was performed using \textsc{enterprise}~\cite{enterprise}, and posterior sampling was performed using \textsc{ptmcmcsampler}~\cite{ptmcmcsampler}, utilising parallel tempering to explore high-dimensional parameter space. 
We performed population synthesis calculations using~\textsc{holodeck}~\cite{ng15pop}.
Corner plots were generated using \textsc{chainconsumer} \cite{Hinton2016}, and PTA sensitivity curves were computed using \textsc{hasasia} \cite{Hazboun2019Hasasia}.

\section*{Acknowledgments}

We thank Rinat Kagirov for assistance with simulating the data and noise.

\bibliography{mybib,collab,soft,book}{}
\bibliographystyle{apsrev4-2}

\appendix
\section{\label{app:tables}Tables with credible levels}

In this section, we provide numerical values for the inferred values of $H_0$ and $\log_{10}\mathcal{M}$ from our simulated data. 
In Table~\ref{tab:time_constraints}, we provide the median values with $1\sigma$ credible intervals for a set of simulations with increasing observational spans. 
In Table~\ref{tab:pulsar_constraints}, we report the median values with $1\sigma$ credible intervals for a set of simulations with a number of CPTA pulsars used increasing from 10 to 40. 
In Table~\ref{tab:source_constraints}, we provide the median values with $1\sigma$ credible intervals for a set of simulations with an increasing number of simultaneously observed CGW sources (P1, P2, P3) with the 20 best CPTA pulsars observed over the period of 40 years.

\begin{table}[H]
\centering
\caption{Median and $1\sigma$ credible levels obtained in parameter estimation with a single simulated CGW source (B1) in the CPTA data. Results are shown for the 40 best pulsars.
The inferred precision of $H_0$ improves by a factor of two as the observation span is increased from 5 to 40 years. 
}
\label{tab:posteriors}
\renewcommand{\arraystretch}{1.4} 
\setlength{\tabcolsep}{12pt}      

\begin{tabular}{|c|c|c|}
\hline
\textbf{Observation Span} & \multicolumn{2}{c|}{\textbf{Estimates}} \\\hline
\textbf{(in years)} & $\log_{10}\mathcal{M}~[M_\odot]$ & $H_0~[\frac{\text{km}}{\text{s}~\text{Mpc}}]$ \\\hline
5  & $9.567^{+0.008}_{-0.006}$ & $68.591^{+4.072}_{-3.503}$ \\\hline
10 & $9.570^{+0.003}_{-0.003}$ & $68.505^{+2.711}_{-2.472}$\\\hline
20 & $9.569^{+0.002}_{-0.001}$ & $66.873^{+2.205}_{-2.396}$\\\hline
30 & $9.568^{+0.001}_{-0.001}$ & $67.131^{+2.162}_{-2.159}$\\\hline
40 & $9.566^{+0.001}_{-0.001}$ & $65.330^{+2.086}_{-2.111}$\\\hline
\end{tabular}
\label{tab:time_constraints}
\end{table}

\begin{table}[H]
\centering
\caption{Median and $1\sigma$ credible levels obtained in parameter estimation with a single CGW source (B1) in the CPTA data.
Results are shown for an observation span of 40 years, for a different number of best pulsars included in the analysis. 
}
\label{tab:posteriors}

\renewcommand{\arraystretch}{1.4} 
\setlength{\tabcolsep}{12pt}      

\begin{tabular}{|c|c|c|}
\hline
\textbf{Array size} & \multicolumn{2}{c|}{\textbf{Estimates}} \\\hline
(\textbf{No. of Pulsars}) & $\log_{10}(\mathcal{M}~[M_\odot])$ & $H_0~[\frac{\text{km}}{\text{s}~\text{Mpc}}]$ \\
\hline
10 & $9.569^{+0.002}_{-0.003}$ & $64.829^{+4.920}_{-5.421}$  \\\hline
20 & $9.571^{+0.002}_{-0.002}$ & $65.146^{+3.054}_{-3.209}$  \\\hline
30 & $9.566^{+0.001}_{-0.001}$ & $63.318^{+2.498}_{-2.472}$  \\\hline
40 & $9.566^{+0.001}_{-0.001}$ & $65.330^{+2.086}_{-2.111}$  \\\hline
\end{tabular}
\label{tab:pulsar_constraints}
\end{table}

\begin{table}[H]
\centering
\caption{Median and $1\sigma$ credible levels obtained in parameter estimation with multiple CGW sources in the CPTA data.
Here, we simulate a 40-year observation with the 20 best CPTA pulsars. 
For the case of one CGW, we simulate P1. 
For the case of two CGWs, we simulate P1 and P2. 
And for the case of three CGWs, we add P3 to the simulation. 
}
\label{tab:posteriors}

\renewcommand{\arraystretch}{1.4} 
\setlength{\tabcolsep}{12pt}      

\begin{tabular}{|c|c|}
\hline
\textbf{No. of CGWs} & \textbf{$H_0~[\frac{\text{km}}{\text{s}~\text{Mpc}}]$} \\
\hline
1 & $52.854^{+12.515}_{-10.071}$ \\\hline
2 & $72.757^{+5.281}_{-5.738}$ \\\hline
3 & $65.051^{+6.628}_{-5.595}$ \\\hline
\end{tabular}
\label{tab:source_constraints}
\end{table}

\section{\label{app:skymap}Positions of pulsars and GW sources}

\begin{figure}
    \centering
    \begin{subfigure}[t]{0.48\textwidth}
        \centering
        \includegraphics[width=\linewidth]{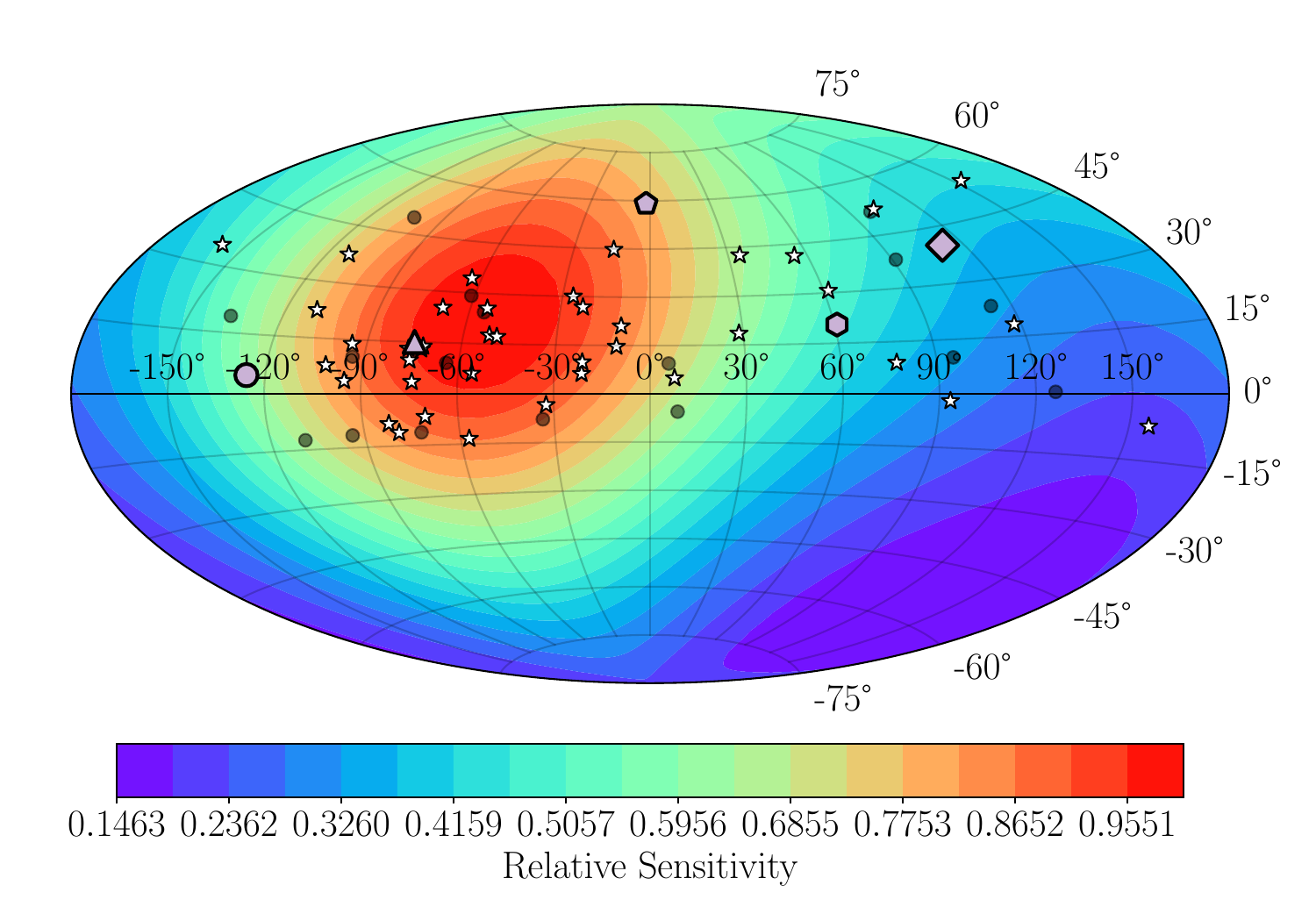}
    \end{subfigure}
    \caption{\justifying Positions and relative sensitivity of the best 40 pulsars (stars) out of 57 in the Chinese Pulsar Timing Array (CPTA).
    The remaining 17 pulsars, not used in the analysis, are shown as black dots. 
    The heatmap displays the relative sensitivity to CGW across the celestial sphere from B1, B2, P1, P2, and P3 binaries represented by diamond, circle, triangle, hexagon, and pentagon, respectively.
    Sensitivity is calculated as the sum of the squared antenna pattern functions, $\sum (F_+^2 + F_\times^2)$, normalised to the maximum value.}
    \label{fig:hasasia}
\end{figure}

In Figure~\ref{fig:hasasia}, we show the relative sensitivity of the CPTA pulsars, where the colour represents values 
\begin{equation}
S(\theta, \phi) = \sum_\text{k=1}^{N_\text{psr}} \left( F_+(\hat{\Omega})^2 + F_\times(\hat{\Omega})^2 \right).
\end{equation}
Here, $S(\theta, \phi)$ is the total sky sensitivity at a particular location $(\theta, \phi)$, $N_\text{psr}$ is the array size (40 in this case). 
The values $F_+$ and $F_\times$ represent the antenna pattern response functions for the ``plus'' and ``cross'' CGW polarisations, respectively. Sensitivity value is normalised to the maximum sensitivity found on the grid to generate the ``Relative Sensitivity colour map.

Pulsar distances are needed when modelling the pulsar term in the CGW signal. In Table \ref{tab:cpta_pulsars} we present the distances and respective uncertainties of the simulated CPTA pulsars used in this work.

\begin{table*}
\centering
\caption{Assumed distances of simulated CPTA pulsars used in this work.}
\label{tab:cpta_pulsars}
\renewcommand{\arraystretch}{1.4}
\setlength{\tabcolsep}{12pt}
\begin{tabular}{|c|c|c|c|}
\hline
\textbf{PSR} & \textbf{Distance (kpc)} & \textbf{PSR} & \textbf{Distance (kpc)} \\
\hline
J0030+0451 & $0.28 \pm 0.10$ & J1857+0943 & $0.90 \pm 0.20$ \\\hline
J0154+1833 & $1.00 \pm 0.20$ & J1903+0327 & $1.00 \pm 0.20$ \\\hline
J0218+4232 & $1.00 \pm 0.20$ & J1910+1256 & $1.95 \pm 0.39$ \\\hline
J0340+4130 & $1.00 \pm 0.20$ & J1911+1347 & $1.00 \pm 0.20$ \\\hline
J0406+3039 & $1.00 \pm 0.20$ & J1918-0642 & $1.40 \pm 0.28$ \\\hline
J0509+0856 & $1.00 \pm 0.20$ & J1923+2515 & $1.00 \pm 0.20$ \\\hline
J0613-0200 & $0.90 \pm 0.40$ & J1946+3417 & $1.00 \pm 0.20$ \\\hline
J0645+5158 & $1.00 \pm 0.20$ & J2010-1323 & $1.00 \pm 0.20$ \\\hline
J0751+1807 & $1.00 \pm 0.20$ & J2017+0603 & $1.00 \pm 0.20$ \\\hline
J1012+5307 & $0.70 \pm 0.20$ & J2022+2534 & $1.00 \pm 0.20$ \\\hline
J1024-0719 & $0.49 \pm 0.12$ & J2033+1734 & $1.00 \pm 0.20$ \\\hline
J1327+3423 & $1.00 \pm 0.20$ & J2043+1711 & $1.00 \pm 0.20$ \\\hline
J1630+3734 & $1.00 \pm 0.20$ & J2150-0326 & $1.00 \pm 0.20$ \\\hline
J1640+2224 & $1.19 \pm 0.24$ & J2214+3000 & $1.00 \pm 0.20$ \\\hline
J1713+0747 & $1.05 \pm 0.06$ & J2229+2643 & $1.00 \pm 0.20$ \\\hline
J1738+0333 & $1.00 \pm 0.20$ & J2234+0611 & $1.00 \pm 0.20$ \\\hline
J1741+1351 & $1.00 \pm 0.20$ & J2234+0944 & $1.00 \pm 0.20$ \\\hline
J1832-0836 & $1.00 \pm 0.20$ & J2302+4442 & $1.00 \pm 0.20$ \\\hline
J1843-1113 & $1.00 \pm 0.20$ & J2317+1439 & $1.89 \pm 0.38$ \\\hline
J1853+1303 & $1.00 \pm 0.20$ & J2322+2057 & $1.00 \pm 0.20$ \\\hline
\end{tabular}
\end{table*}

\section{\label{app:snr_v_time} SNR for CGW sources}
The SNR values for the search of continuous gravitational waves (CGWs) from the binaries listed in Table \ref{tab:injection}, using an array of the 40 best CPTA pulsars, are presented in Table \ref{tab:SNR_values} for different observation spans.

\begin{table*}
\centering
\caption{\justifying Analytical all-sky SNR ($\rho$) for CGW from the sources listed in Table \ref{tab:injection} observed using an array of 40 best CPTA pulsars at different observation spans.}
\label{tab:SNR_values}
\renewcommand{\arraystretch}{1.4}
\setlength{\tabcolsep}{12pt}
\begin{tabular}{|c|c|c|c|c|c|}
\hline
\textbf{CGW Source} & \multicolumn{5}{c|}{\textbf{SNR ($\rho$)}} \\\hline
 & \textbf{5 yr} & \textbf{10 yr} & \textbf{20 yr} & \textbf{30 yr} & \textbf{40 yr} \\
\hline
B1 & 36.169 & 54.149 & 77.344  & 95.642  & 110.797    \\\hline
B2 & 13.873 & 20.551 & 28.933  & 35.620  & 41.336     \\\hline
P1 & 13.750 & 68.375 & 108.300 & 132.279 & 153.493    \\\hline
P2 & 1.257  & 1.961  & 3.331   & 4.162   & 4.792      \\\hline
P3 & 0.613  & 0.865  & 1.272   & 1.544   & 1.810      \\\hline
\end{tabular}
\end{table*}

\section{\label{app:full_pe}Parameter estimation results}
\begin{figure*}[!t]
    \centering
    \includegraphics[width=0.96\linewidth]{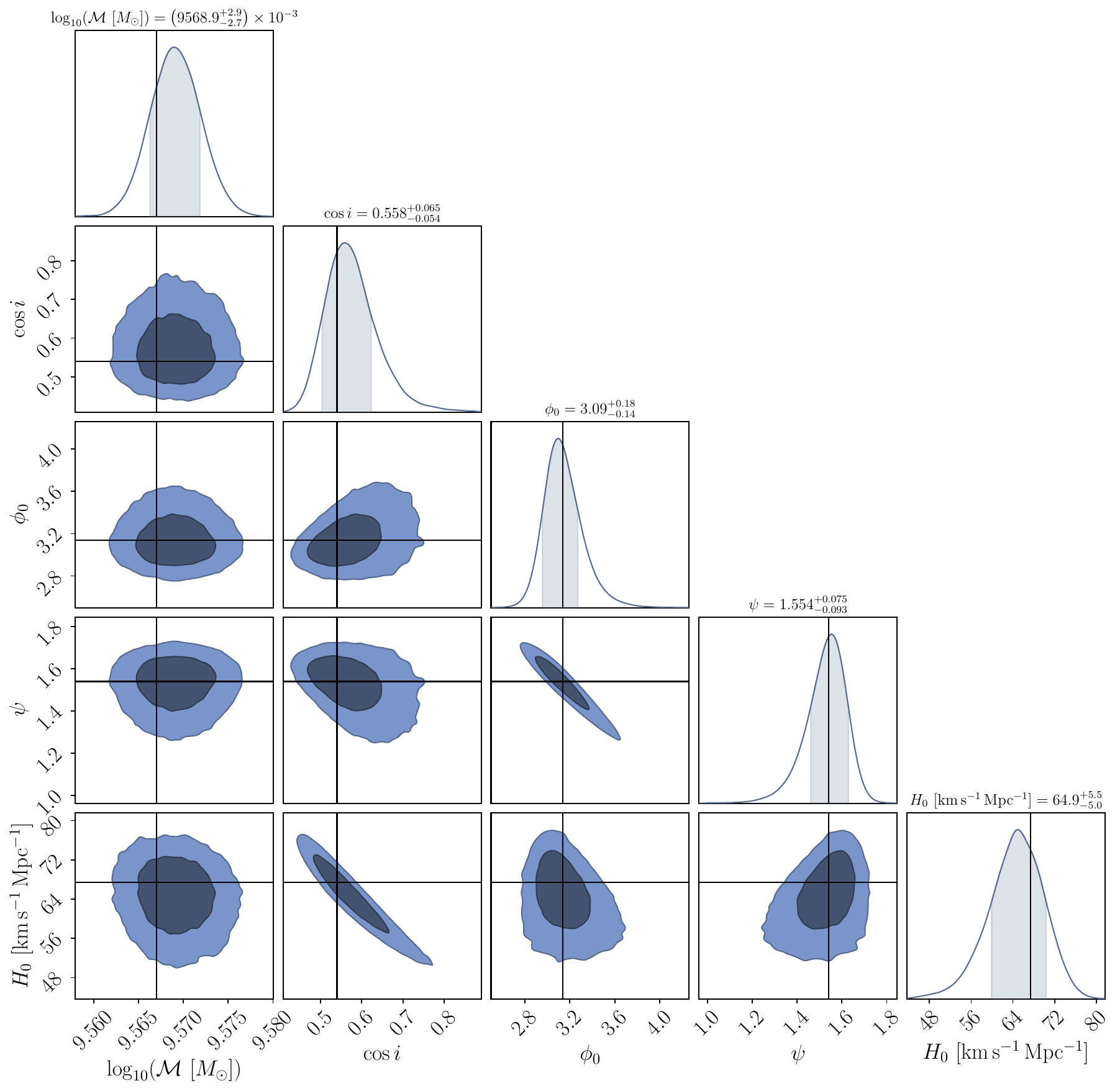}
    \caption{\justifying Posterior probability for the Hubble constant ($H_0$) and CGW parameters in a simulated targeted search for B1, an SMBHB target. The synthetic data represents the 10 best CPTA pulsars over a 40-year observation span. The diagonal histograms show the marginalised posteriors for each parameter, the shaded region indicates $1\sigma$ credible region with the simulated truth values represented with black lines. Off-diagonal panels correspond to the posteriors for parameter pairs with the shaded regions showing credible $1$ and $2\sigma$ credible regions.}
    \label{fig:10_40}
\end{figure*}

In the main body of the paper, we mostly discuss the results of parameter estimation of $H_0$. 
In this section, we provide an outline of the estimation of CGW parameters. 
Figure~\ref{fig:10_40} shows the results corresponding to the 10 best CPTA pulsars observed over 40 years with one CGW source injection (B1). 
The diagonal panels are the marginalised posterior probability distributions for the CGW parameters.
The off-diagonal plots show the two-dimensional posteriors for parameter pairs, illustrating correlations between parameters. 
The orange lines are the simulated values, demonstrating that the parameter estimation is accurate. 

\end{document}